\documentstyle[prl,aps,amssymb,graphicx,twocolumn]{revtex}
\def\K{{\cal K}}
\def\r{\vec{r}}
\def\w{\omega}
\def\wc{\w_{\text{c}}}
\def\const{\text{const}}
\begin{document}


  \draft
\title{Reply to Comment by A. Moroz}
\author{Sajeev John and Ovidiu Toader}
\address{Department of Physics, University of Toronto, \\
60 St. George Street, Toronto, Ontario, Canada M5S-1A7}
\date{January 19, 2000}
\maketitle

In his comment, Moroz questions the validity of the near band edge
(effective mass) approximation to the total photon density of states
(DOS) as a useful representation of the local density of states (LDOS)
experienced by a single radiating atom or molecule located at a
particular position $\r$ within a photonic crystal (PC). In this
approximation, the band edge DOS takes the form:
$$
\rho(\w) \approx \const |\w-\wc|^\eta
$$
where $\eta=-0.5$ for a 1-d PC and $\eta=0.5$ for a 3-d PC.  We
reassert that this behaviour indeed applies to the LDOS as well as the
DOS. However, the frequency range over which this behaviour is
realized depends sensitively on $\r$. In particular, if $\r$ is chosen
near a node of the electromagnetic field intensity
$\left|\vec{E}(\vec{r})\right|^2$, then $\w$ must be chosen very close
to $\wc$ before the asymptotic behaviour is realized. The seemingly
arbitrary exponents obtained by Moroz are simply an artifact of
fitting the asymptotic form to numerical data for a frequency $\w$
which is not sufficiently close to $\wc$ at certain positions $\r$.

We consider precisely the example quoted by Moroz in his comment and
assume that the LDOS has the asymptotic form:
$$
\rho(\w,\r) = \K(\r) |\wc - \w|^\eta
$$
Near the lower band edge of the first photonic band gap
($\w\lessapprox \wc$) we define $u \equiv 1 - {\w\over\wc} > 0$.
In order to numerically estimate the exponent $\eta$, we write:
$$
y \equiv \log_{10} \rho = \eta (\log_{10} u + \log_{10} \wc) +
\log_{10} \K(\r)
$$
Using equations (4) and (7) of Moroz's paper \cite{Moroz_1} we plot 
(in Fig. 1a) $y$ as a function of $z\equiv\log_{10}u$ for 8
different positions $\r$ in the 1-d unit cell of the example quoted in
the above comment. The asymptotic behaviour of $dy/dz$ for large
negative values of $z$ ($\w\rightarrow\wc$) yields the exponent $\eta$
(see Fig. 1b). In this model the lower band edge mode intensity
vanishes at $x\equiv|\r|=0.5$ (center of air region) and has a maximum
at $x=0.0$ (center of dielectric slab). For all cases the asymptotic
behaviour ($\w\rightarrow\wc$) yields the common exponent
$\eta=-0.5$. However arbitrary values of $dy/dz$, and hence $\eta$, may
be erroneously inferred by choosing too large a value of
$|\w-\wc|$. This is particularly evident near the node of 
 the field intensity.

We conclude that although the LDOS is sensitive to the actual position
$\r$, the exponent $\eta$ is indeed universal except on a set of
measure zero, namely the field intensity nodes. The seemingly
arbitrary exponents quoted by Moroz are somewhat misleading. On the
other hand, inhomogeneous line broadening is a very important
and relevant ingredient which must be incorporated into theoretical
models which aim to interpret experiments involving a distribution of
atoms in a PC.

\begin{figure}
\centerline{\includegraphics[height=4in,width=3.1in]{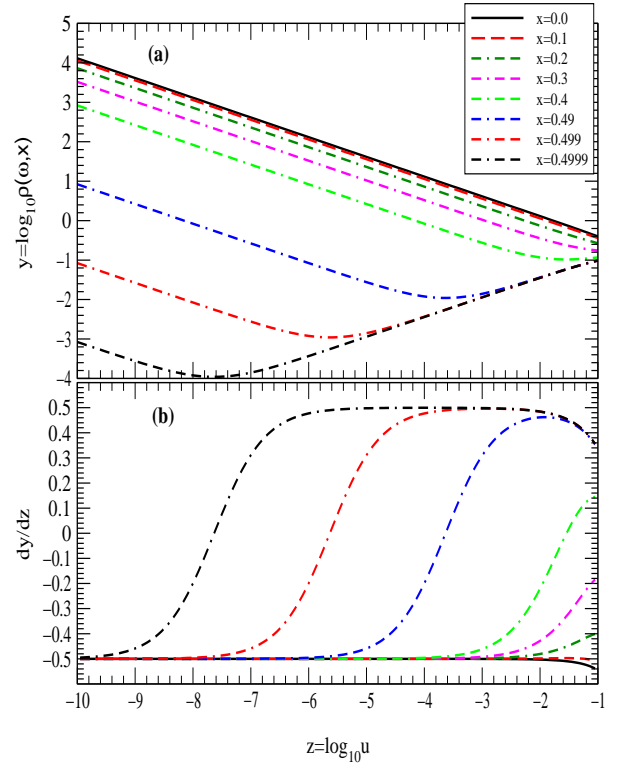}}
\caption{Top picture shows $\log_{10} \rho\left(\w,x\right)$
as a function of $z=\log_{10}u$ for 8 positions in the unit cell. Here
$x=|\r|$. Bottom
picture shows the slope of the curves. In all cases the asymptotic
behaviour ($\w\rightarrow\wc$) yields $\eta=-0.5$.}
\label{fig:1}
\end{figure}

\bibliography{PC.bib,CP.bib,books.bib}
\bibliographystyle{prsty}

\end{document}